# Sun-tracking optical element realized using thermally activated transparency-switching material

Harry Apostoleris,[1] Marco Stefancich,[3] Samuele Lilliu[2] and Matteo Chiesa*[1]

[1]*Laboratory for Energy and Nanoscience, Masdar Institute of Science and Technology, Masdar City, Abu Dhabi, UAE*
[2] *Masdar Organic Photovoltaics Laboratory, Masdar Institute of Science and Technology, Masdar City, Abu Dhabi, UAE*
[3] *National Research Council, IMEM, Parco Area delle Scienze 37/A - 43124 Parma, Italy*
* *mchiesa@masdar.ac.ae*

We present a proof of concept demonstration of a novel optical element: a light-responsive aperture that can track a moving light beam. The element is created using a thermally-activated transparency-switching material composed of paraffin wax and polydimethylsiloxane (PDMS). Illumination of the material with a focused beam causes the formation of a localized transparency at the focal spot location, due to local heating caused by absorption of a portion of the incident light. An application is proposed in a new design for a self-tracking solar collector.

1. **Introduction**

Sun tracking is a critical issue in the development of concentrated photovoltaics (CPV) as it is necessary for all but low concentrations [1], and therefore has a great impact on efforts to promote further CPV penetration. Concentration may enable more cost effective solar energy conversion by allowing reduced use of photovoltaic material, or the inclusion of more efficient cells with reduced area, coupled with cheap optical elements. This potential has been realized, in the extreme, through the use of high-concentration collectors incorporating multijunction solar cells to achieve record efficiencies, with the present record standing at 36.7% module efficiency [2], with a record cell efficiency of 44.7% [3]. However, the potential of concentrating systems is limited by the requirement of sun tracking, which increases cost and creates a physical encumbrance making CPV unsuitable for many small-scale applications. The origin of the tracking requirement is the thermodynamic concentration limit arising from the principle of étendue conservation and given by $C \leq n^2 / \sin^2 \theta_{acc}$ for a device with refractive index $n$ and accepting light from a cone with half-angle extent $\theta_{acc}$, providing a concentration $C$ [4]. As a consequence, increasing the concentration of a solar collector necessarily reduces the acceptance angle $\theta_{acc}$ and therefore the timespan during which the solar disk can be maintained within the acceptance cone of a stationary concentrator. At present, tracking is accomplished by mechanically rotating the collector to maintain its orientation towards the sun over the course of the day. Depending on the precision required, this may be accomplished using a single-axis or double-axis mechanical tracker [5], with increased precision generally requiring greater physical size and cost. However there is no physical requirement that this 'rotational' approach to sun tracking is the only possible strategy. For example, it has been demonstrated that with appropriate optical design, tracking over a wide angular range may be achieved via small lateral translations of a collecting element [6, 7], or by incorporating an optically active element which can vary its optical properties in response to the changing solar angle to create a 'self tracking' system that can track the sun reactively with no external inputs [8, 9].

It has been proposed that tracking may be achieved via a localized, light-activated change in transparency to create a concentrating light trap with a dynamic aperture that moves in response to the changing solar angle to admit light over a wide angular range (Fig. 1(a)) [10]. This is achieved by initially focusing light with a lens or set of lenses onto the receiving surface of a secondary concentrator, which is covered by a transparency-switching material; the local transparency formed as a result of this illumination then follows the focal spot as it moves across the surface due to the changing solar angle. Progress in developing this design has to date been limited mainly by the lack of an appropriate material from which to create the moving aperture element. Recently, a material satisfying many of the requirements for such an application has been identified in the form of a thermally responsive transparency-switching composite of paraffin wax and polydimethylsiloxane (PDMS)[11]. The material undergoes a shift at the melting point of the paraffin between a high-temperature transparent ('on') state and an low temperature opaque ('off') state in which light is strongly scattered. The mechanism of the switching has been identified as a transformation of the microstructure caused by the reversible formation and destruction of dispersed wax crystals, which act as scattering centers for light, in the PDMS matrix resulting from the paraffin phase transition [12] (Fig. 1(b)). In the present work, we make use of this behavior to create a light-tracking aperture suitable for the desired application.

## 2. Theory

The potential value of this design can be demonstrated in principle by an idealized theoretical description of its behavior as a concentrator. For this theoretical discussion we can consider an ideal nonimaging concentrator, which could be approached in real life by, for example, a CPC. To provide a numerical result, we consider a concentrator with an acceptance angle $\theta_{acc} = 10°$, containing a solar cell at its exit aperture. For an ideal concentrator, the concentration of rays within the acceptance cone is equal to the thermodynamic limit $1/\sin^2\theta_{acc}$ (for a hollow concentrator), which in this case is 33x. In Fig. 2(a) the same concentrator is used in the configuration of ref.[10], with a transparency-switching covering over the entrance aperture and a primary optic to focus light onto this covering. Prior work by various authors provides sufficient information to estimate the performance of this ideal system analytically. For the primary optic, we consider a configuration described by Zagolla et al. [13]. This system, composed of two aspheric lenses with diameter 25 mm, was found to be effective at concentrating light onto a flat image surface, whose width is the same as the lens diameter, over a ±23° range of incidence angles, defining the tracking range of the system. The spot diameter was shown to be less than 1 mm over the whole tracking range. Based on the presented results of ref. [13], and our own corroborating ray trace simulations on a similar system, the focused rays have an angular divergence $\theta_{in} \approx \pm 30°$.

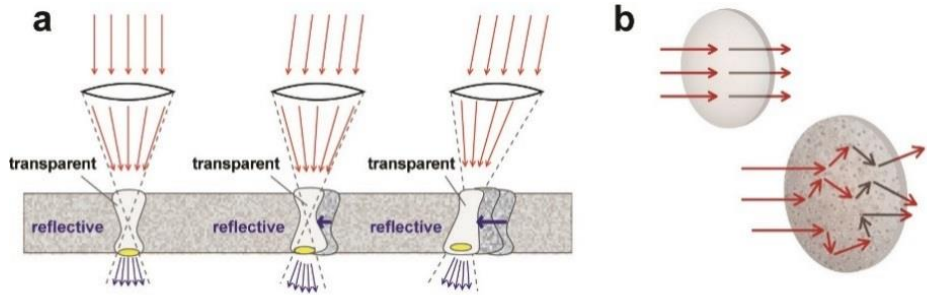

**Figure 1**. a) A proposal to use a localized, light-activated transparency shift to create a light-tracking aperture in b) a thermally activated transparency-switching paraffin-PDMS composite. At low temperatures the composite becomes inhomogeneous and strongly scattering due to the phase change of the paraffin.

Using this information, the optical efficiency of the concentrator can be estimated using a statistical model which has previously been described by us and others [14, 15] in the context of a similar light-trapping system [16]. We can consider the light absorbed by the solar cell in this way: when light initially enters the concentrator, some fraction is absorbed immediately and the remainder is rejected. This initial absorption can be given by $\alpha_{init} \approx \sin^2\theta_{acc} / \sin^2\theta_{in}$ if the incoming light is considered to be diffuse within the cone defined by $\theta_{in}$. Due to the reflective covering on the entry aperture, most of these rejected are reflected back into the concentrator, with a diffuse profile. On all subsequent passes, the probability of absorption per pass may then be given by $\alpha_{diff} = \sin^2\theta_{acc}$ and the probability of escape by the ratio

$p_{esc} = d_{aperture}^2 / d_{entrance}^2$, where $d_{aperture}$ and $d_{entrance}$ are the diameters of the small input aperture and the whole concentrator entrance surface, respectively. This holds in the ideal case where transmission through the aperture is 100% and through the rest of the entrance surface is 0%. Carrying out the summation over all possible passes, as in refs. [15] and [14], leads ultimately to the expression

$$\eta = \alpha_{init} + (1 - \alpha_{init})(1 - p_{esc}) \frac{\alpha_{diff}}{1 - (1 - \alpha_{diff})(1 - p_{esc})}. \tag{1}$$

Using the parameter values listed in this section, our ideal system can, according to Eq.1, achieve an optical efficiency of up to 95% (concentration of ~31 suns, red line in Figure 2c). The efficiency is strongly dependent on the escape probability. This efficiency may be maintained over the full ±23° tracking range. A figure of merit for the tracker can be the ratio of the integrals of the concentration over the incidence angle; in our idealized case, this is simply $C_{tracking}\theta_{tracking} / C_{nontracking}\theta_{nontracking} = 31 \times 0.40 / 33 \times 0.17 = 2.2$, representing a 2.2x enhancement in concentration over the whole angular range for this geometry relative to the conventional concentrator.

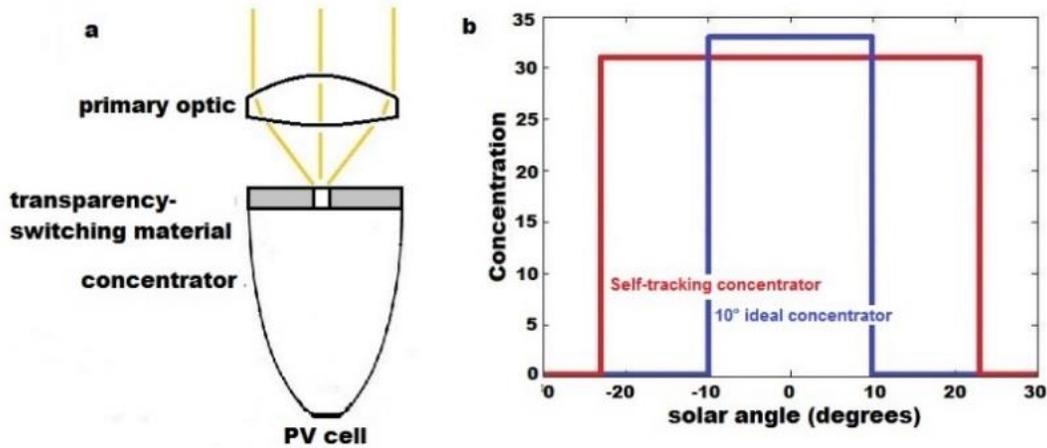

**Figure 2**. a) schematic of the self-tracking system consiting of a lens, a transparency-switching layer and a compound parabolic concentrator to guide light onto a PV cell; b) flux concentration as a function of solar angle for an ideal non-tracking concentrator (blue line) and an idealized self-tracking concentrator of the type shown in (a).

## 3. Experiment

The material is prepared by mixing paraffin wax (Sigma Aldrich product 76228) and uncured PDMS (Dow Chemical SYLGARD 184) in a 1:9 ratio. To enable absorption, a small amount (0.1 wt %) of carbon black paint (Winton Oil Colour 24/Ivory Black from Winsor & Newton) is added to the material. The standard Sylgard curing agent is added to in a ratio of 1:10 and the mixture is stirred manually until all components are fully distributed. The gel-like substance is poured into circular PLA molds of 20 mm radius fixed to a polycarbonate substrate and cured by heating in an oven for 2 hours at 90°C. The cured samples are solid rubbery disks which can be peeled off of the substrate. They undergo transparency switching at about 45°C, which is the manufacturer-provided melting temperature of the paraffin. As has been established in previous work by us and others [11, 12], the transition temperature of the composite can be controlled by using paraffins with different melting points.

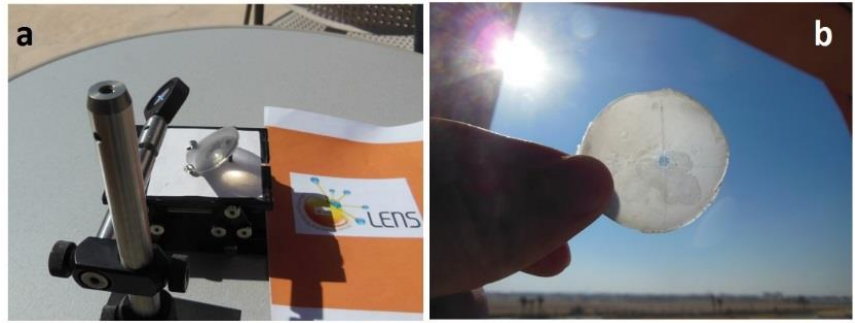

**Figure 3**. a) outdoor setup for light-tracking demonstration; b) localized transparency in the material resulting from illumination with focused sunlight.

The sun tracking demonstration was set up on the campus of the Masdar Institute in Abu Dhabi. A sample of the material ~1.5 mm thick was mounted on a stage at an angle facing the sun at noon. To prevent thermal conduction between the stage and the material, the sample was in contact with the stage only at the bottom edge and three support points. A converging lens (d=25.4 mm, f=50 mm) was placed parallel to the sample in line with the sun, with the sample at the focal point of the lens, creating a focal spot near the center of the sample (Fig. 3(a)). To document the tracking behavior, two sets of time lapse photographs were taken of the sample at a rate of 1/minute for one hour, one set with the lens covered to show more clearly the location of the transparency, and the other with full illumination to demonstrate that the light passes through the aperture unscattered.

## 4. Results

After the sample is positioned at the focal point of the lens, a transparent aperture forms at the spot location on a timescale of about 10s. The aperture has a clearly defined border and features can be seen clearly through it. If the sample is illuminated for only a short period of time (in this sample about one minute), the transparency remains highly localized around the focal spot location, as seen in Fig. 3(b).

If the lens is removed following the aperture formation, switching back to the off state will occur quickly and the aperture will close. This is illustrated in Fig. 4. From the light spot on the white backing, one can observe the transition from the fully opaque state, characterized by the partial transmission of diffused light, to the locally transparent state with full transmission through the aperture (Fig. 4(a-d)). After the lens is removed, a 'pinhole' spot created by transmission of direct sunlight through the aperture can be seen on the backing, gradually vanishing as the aperture closes (Fig. 4(e-h)).

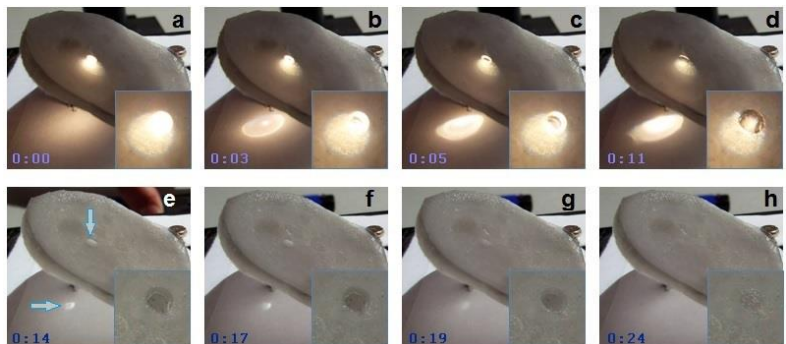

**Figure 4**. Formation and closing of the transparent aperture under solar illumination. a-d) observation of the white backing in the large figures shows transition from partial diffuse transmission (a) to full transmission forming a well-defined spot (d). Insets show formation of local transparency. e-h) Loss of transparency after removal of the lens. Transmission of unconcentrated sunlight through the aperture forms the 'pinhole' projection visible in e-g.

When the lens is left in place, over the first several minutes of illumination the aperture increases in size before stabilizing at a diameter of approximately 5 mm as the equilibrium is reached between absorptive thermal inputs and conductive losses into the surrounding material. During the following hour, the solar angle changes sufficiently that the focal spot moves from the center of the sample almost to its edge, as seen in Figure 4. The aperture position changes with the spot location, remaining centered on the focal point for the duration of the demonstration (Figure 5, top frames). The aperture size remains constant, with strong localization around the focal point (Figure 5, bottom frames).

## 5. Discussion

The results of this demonstration serve as a proof-of-concept for a light-responsive aperture based on a thermally activated transparency switching material. While other uses may be found, we focus in this discussion on the application for which the concept was originally developed, that of a concentrating light trap with a moving aperture. The purpose of the material in this case is to admit light in concentrated form through a small aperture, then confine that light inside the trap by internally reflecting any rays on an escape trajectory. In the configuration proposed by Stefancich et al. in ref. [10], or indeed in any light-trapping configuration, the performance depends on limiting the optical losses associated with those reflections. These losses may come from absorption of light by the material; by transmission through the material to the outside; or by rays escaping through the aperture. Optimization of the material is focused on minimizing these three sources of loss. Absorptive losses result from absorption of trapped light by the pigment incorporated into the material. Absorption is essential to the operation of the element and therefore cannot be eliminated absolutely; however, if a selectively absorbing pigment is used, the element may function by using the selected range of wavelengths to activate the tracking element while transmitting other wavelengths without absorption. If photovoltaic cells are being employed, an appropriate pigment would absorb below the bandgap of the utilized PV material, or alternatively the short wavelengths that have a high photon energy and are most responsible for the thermal gains detrimental to the performance of the PV cell. Minimizing transmittance through the 'reflective' regions is equivalent to reducing the off-state transmittance, which is currently being approached with a double strategy of identifying new component materials that maximize the internal refractive index contrast of the composite, and manipulating the crystallization kinetics to increase the scattering center density. Aperture losses are reduced by minimizing the aperture size, which itself is achieved by manipulating the thermal transport behavior so that the equilibrium size of the transparent region is as close as possible to the spot size. This may be achieved by attempting to control the thermal conductivity of the material but is also affected by the degree of pigmentation, which dictates the rate of thermal input. These optimization approaches are the subject of ongoing research.

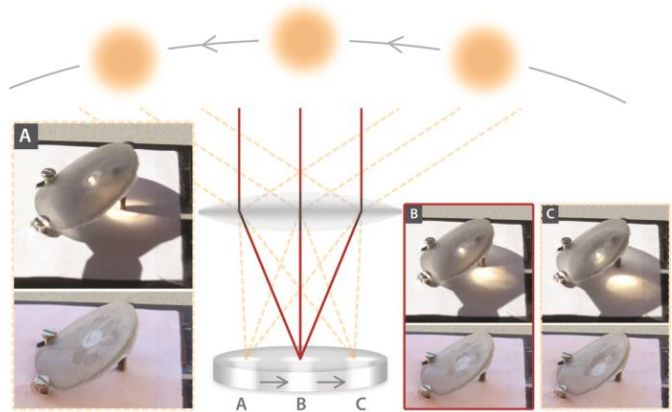

**Figure 5**. The motion of the aperture tracking the solar focal spot over a one-hour period. Top images show the transmitted beam; bottom images taken with the lens blocked show the localization of the transparency throughout the tracking motion.

## 6. Conclusions

We have demonstrated a light-responsive aperture based on a thermally activated transparency-switching material. The material forms a localized transparency in response to illumination by a focused light beam, which moves in response to changes in the beam position so that the transparency is always centered on the focal spot location. The operating principle of the moving-aperture light trap, the application for which this element has been developed, has been summarized and the necessary optimizations and improvements to the element required for this use have been described along with the research strategies being employed to reach each optimization target. In this context, we believe the demonstration presented here represents a significant development in the field of optically active materials, with substantial application potential in solar energy and beyond.